\documentclass[a4paper]{article}
\usepackage{frascatiphys}
\usepackage[pdftex]{color,graphicx}
\begin{document}
\title{
HIDDEN PHOTONS IN BEAM DUMP EXPERIMENTS AND IN CONNECTION WITH DARK MATTER}
\author{
Sarah Andreas        \\
{\em Deutsches Elektronen-Synchrotron DESY, Hamburg, Germany}}
\maketitle
\baselineskip=11.6pt
\begin{abstract}
Hidden sectors with light extra U(1) gauge bosons, so-called hidden photons, recently received much interest as natural feature of beyond standard model scenarios like string theory and SUSY and because of their possible connection to dark matter. This paper presents limits on hidden photons from past electron beam dump experiments including two new limits from experiments at KEK and Orsay. Additionally, various hidden sector models containing both a hidden photon and a dark matter candidate are discussed with respect to their viability and potential signatures in direct detection.
\end{abstract}
\baselineskip=14pt

\section{Introduction}

Gauge bosons of an extra U(1) symmetry in a hidden sector, so-called hidden photons, are well motivated since they arise naturally in string compactifications like the heterotic string or type-II string theories. Since the standard model (SM) is not charged under this new gauge group, there is no direct coupling and the interactions with the visible sector, and consequently the experimental constraints, are very weak. The hidden photon is additionally of great interest as it provides a solution to the discrepancy between the SM prediction of the muon anomalous magnetic moment and its experimentally measured value~\cite{Pospelov:2008zw}. Furthermore, models containing in the hidden sector a dark matter (DM) candidate which interacts with the visible sector via a light hidden photon have attracted much attention in the context of recent astrophysical observations~\cite{Fayet:2007ua:ArkaniHamed:2008qn:Cheung:2009qd,Pospelov:2007mp:Chun:2010ve:Mambrini:2011dw:Hooper:2012cw}.

At low energies, the dominant interaction of the hidden photon $\gamma'$ with the visible sector is through kinetic mixing with the ordinary photon. This can, for example, be generated from loops of heavy particles charged under both U(1)s~\cite{Holdom:1985ag}. Integrating out those particles gives as an estimate for the size of the kinetic mixing the order of a loop factor $\mathcal{O}(10^{-3}-10^{-4})$. We then impose the following relation between the hidden gauge coupling $g_h$ and the kinetic mixing $\chi$
\begin{equation}
\chi = \frac{g_Y g_h}{16 \pi^2} \ \kappa ,\label{eq-kappa}
\end{equation}
where $\kappa$ is $\sim\mathcal{O}(1)$ and depends on the masses of the particles in the loop.

For the most simple hidden sector with just an extra U(1) symmetry and the corresponding hidden photon $\gamma'$, the low energy effective Lagrangian describing the kinetic mixing with the photon is then given by
\begin{equation}
{\cal L} = -\frac{1}{4}F_{\mu \nu}F^{\mu \nu} - \frac{1}{4}X_{\mu \nu}X^{\mu \nu}
- \frac{\chi}{2} X_{\mu \nu} F^{\mu \nu} + \frac{m_{\gamma^{\prime}}^2}{2}  X_{\mu}X^\mu + g_Y j^\mu_{\mathrm{em}} A_\mu,
\end{equation}
where $F_{\mu\nu}$ is the field strength of the ordinary electromagnetic field $A_\mu$ and $X_{\mu\nu}$ is the field strength of the hidden U(1) gauge field $X_\mu$. A mass $m_{\gamma'}$ for the hidden photon can be generated either from the Higgs mechanism or from the St\"{u}ckelberg mechanism. In both cases, masses around the GeV-scale can be obtained naturally but much smaller values are also possible~\cite{Goodsell:2009xc:Cicoli:2011yh}. Masses in the GeV range can be tested and constrained especially by electron beam dump experiments. This has been studied in~\cite{Bjorken:2009mm} for past beam dump experiments at SLAC and Fermilab and further in~\cite{Andreas:2012mt} by taking into account the experimental acceptancies and two additional experiments at KEK and in Orsay. An overview of all current experimental constraints on the hidden photon for the MeV to GeV mass range is given in~\cite{Andreas:2012mt,Andreas:2011xf:Andreas:2012xh}.

In general, the hidden sector can contain not only gauge but also matter fields. The attractive possibility of DM in the hidden sector interacting via a hidden photon has been studied in various models for different ranges of DM and hidden photon masses, in particular, GeV-scale dark forces~\cite{Fayet:2007ua:ArkaniHamed:2008qn:Cheung:2009qd,Pospelov:2007mp:Chun:2010ve:Mambrini:2011dw:Hooper:2012cw,Andreas:2011in,GeVscale} but also massless U(1)s~\cite{MasslessHP}. 

This paper summarizes the current status of limits on hidden photons from electron beam dump experiments in Sec.~\ref{sec-HP}, based on the results presented in~\cite{Andreas:2012mt}. In Sec.~\ref{sec-DM}, a toy-model as well as several supersymmetric models for DM interacting via a hidden photon are discussed regarding the DM relic density and signatures in direct detection experiments, following the analysis of~\cite{Andreas:2011in}.

\section{Hidden photons in electron beam dump experiments}\label{sec-HP}

\begin{figure}[htb!]
\begin{center}
\begin{minipage}{5.3cm}
\vspace{-6cm}\includegraphics[height=3cm]{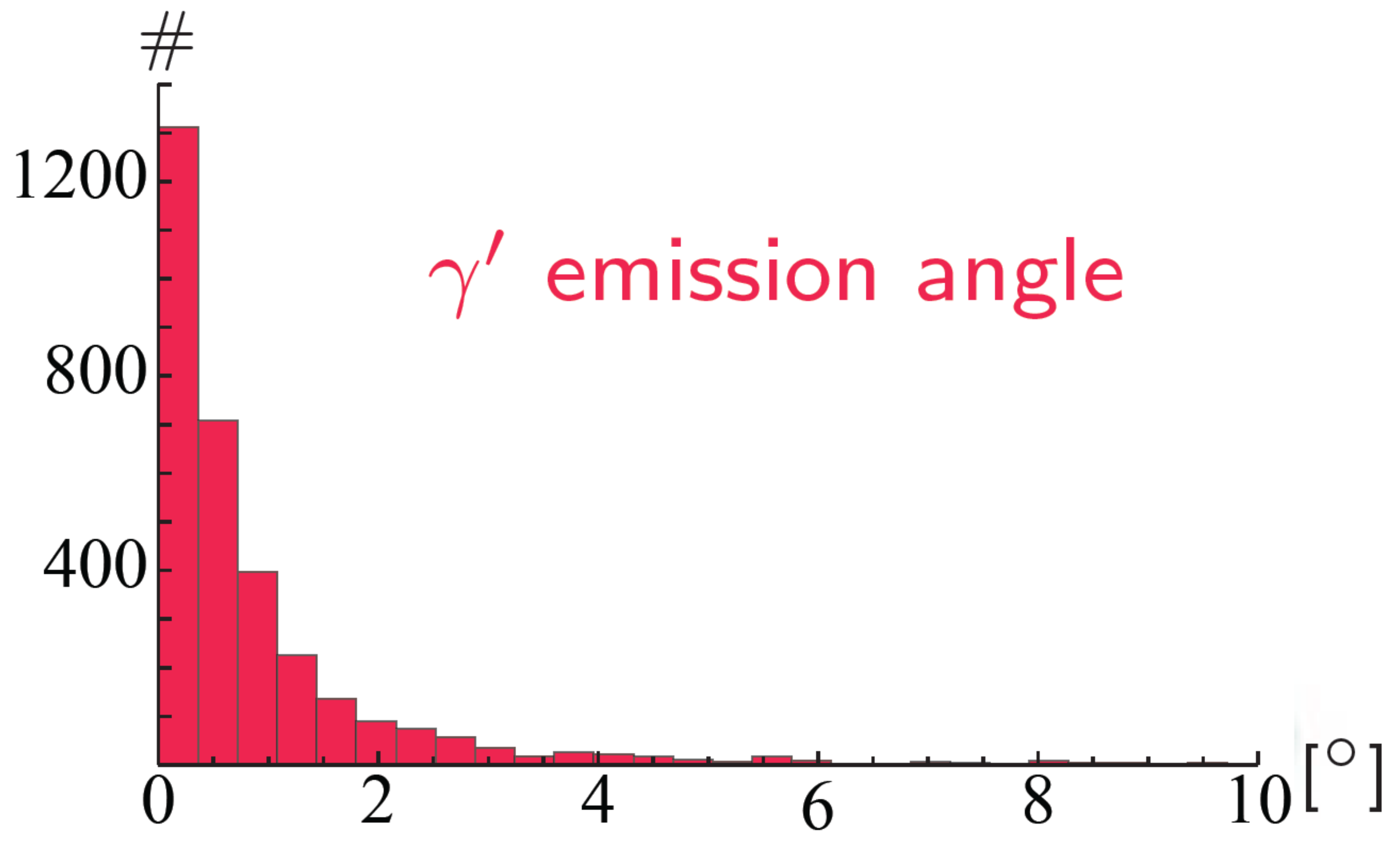}

\includegraphics[height=3cm]{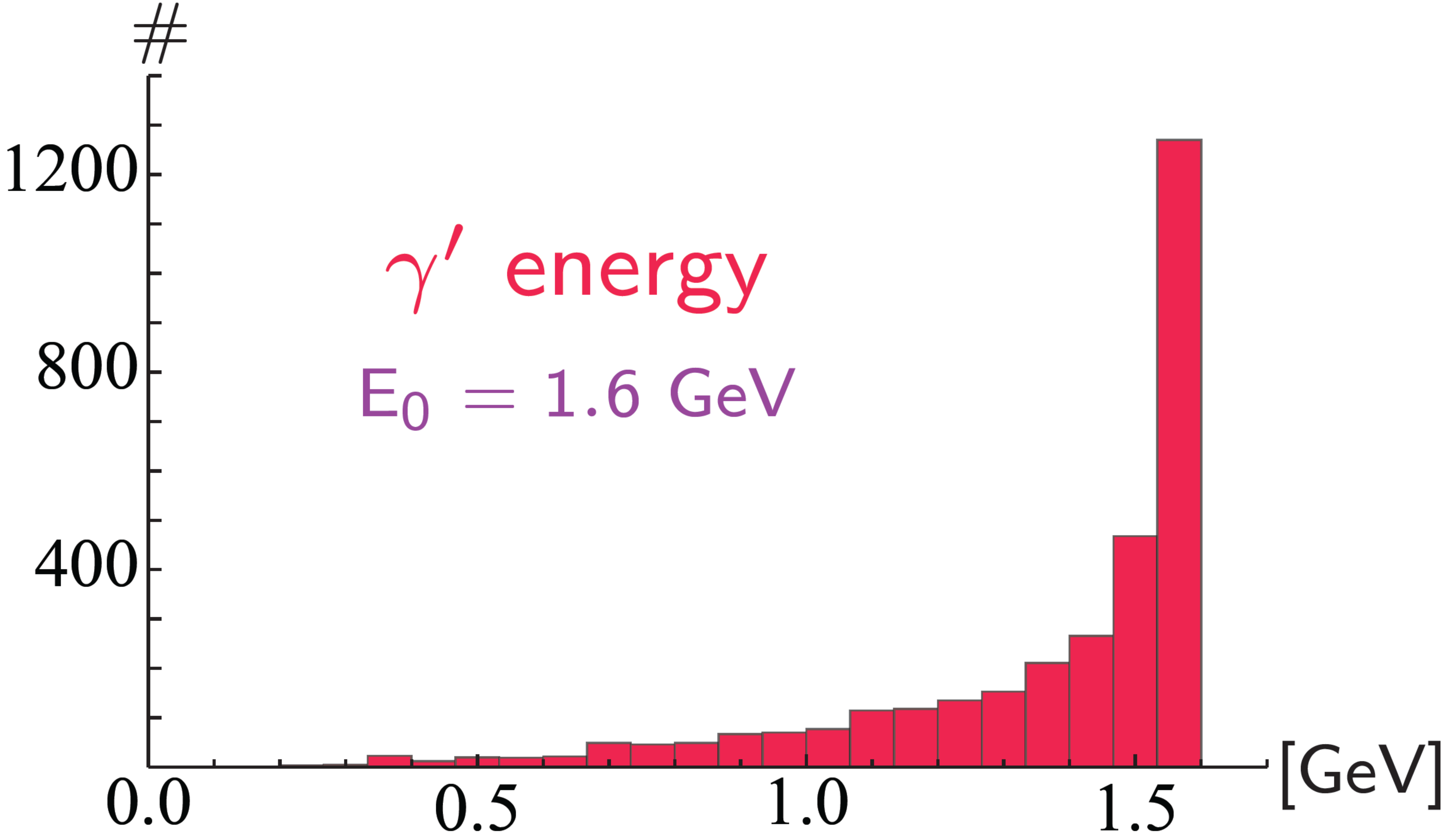}
\end{minipage}
\hspace{0.3cm}\includegraphics[height=6cm]{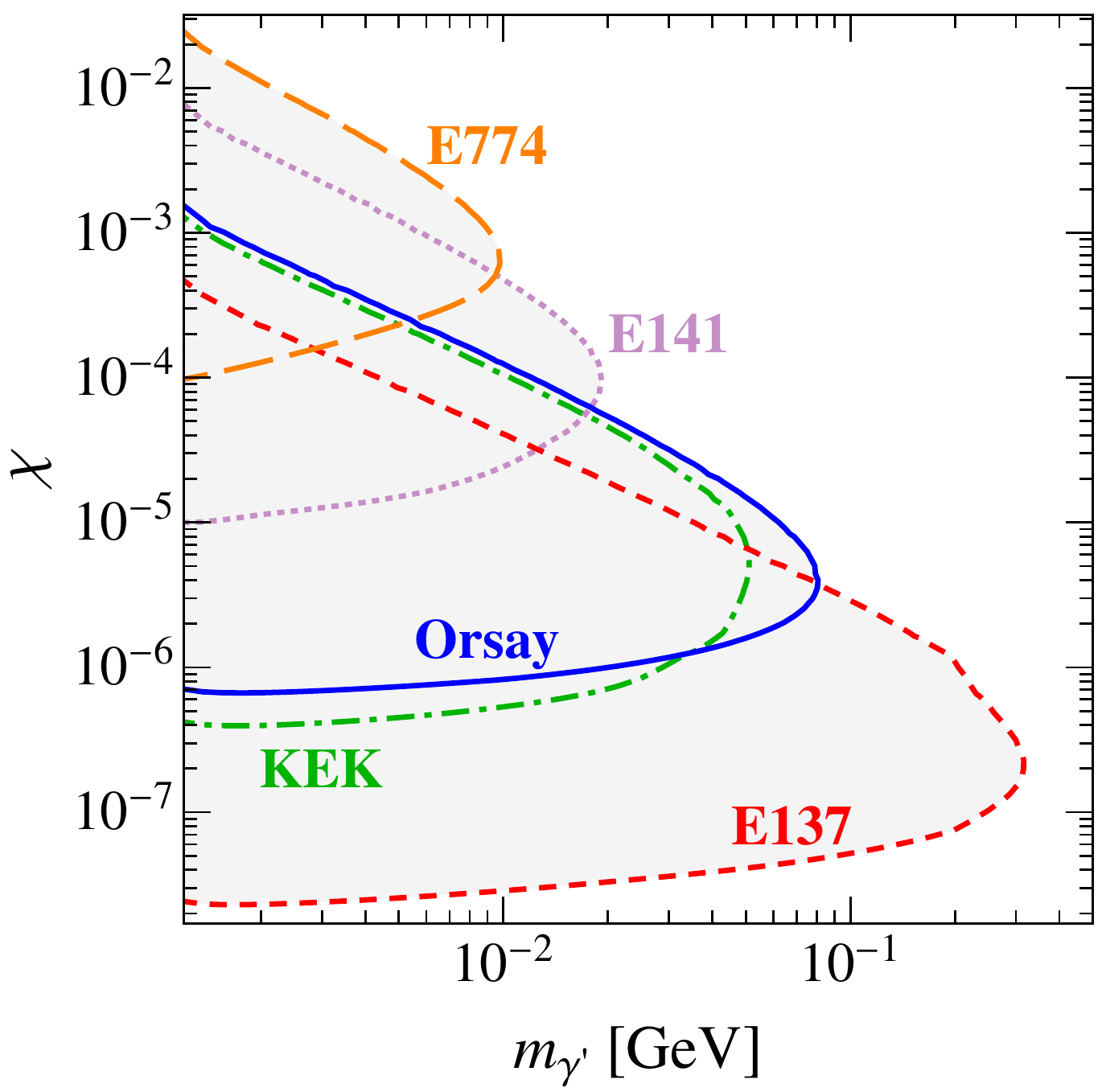}
\caption{\textit{left:} Hidden photon emission angle \textit{(top)} and energy \textit{(bottom)} from Monte Carlo simulations with \textsc{MadGraph} for a beam energy of 1.6 GeV and a total of 3200 hidden photons produced. \textit{right:} Limits on hidden photons from electron beam dump experiments at SLAC (E141, E137), Fermilab (E774), KEK and in Orsay~\cite{Andreas:2012mt}.} \label{fig-HPbeamdump}
\end{center}
\end{figure}

Hidden photons are produced in a process similar to ordinary Bremsstrahlung off an electron beam incident on a target. They are emitted at a small angle in forward direction and carry most of the beam energy, cf. Fig.~\ref{fig-HPbeamdump} (left). Due to their feeble interaction with SM particles they can traverse the dump and be observed in the detector through their decay into fermions, e.g., into $e^+ e^-$.

The production cross section of hidden photons is roughly given by
\begin{equation}
\frac{d\sigma}{dx_e} ~\simeq~ 4 \alpha^3 \chi^2 \ \xi(E_e, m_{\gamma'}, Z, A) \ \sqrt{1 -
\frac{m_{\gamma'}^2}{E_e^2}} \ \frac{1 - x_e + \frac{x_e^2}{3}}{m_{\gamma'}^2
\frac{1-x_e}{x_e} + m_e^2 x_e} \label{eq-dsdxapprme} \ .
\end{equation}
where $x_e=E_{\gamma'}/E_e$ is the fraction of the incoming electron's energy $E_e$ carried by the hidden photon, $m_e$ is the electron mass and $\xi(E_e, m_{\gamma'}, Z, A)$ is an effective flux of photons that takes into account atomic and nuclear form factors and is approximately proportional $\xi \sim Z^2$ for the mass range of interest. 

For the thick target experiments under consideration, the number of hidden photon events observable via the decay products can be estimated as
\begin{equation}
N \simeq  \frac{N_e N_0 X_0}{A} \! \int \!\! dE_{\gamma'}\! \int \!\! dE_e \!\int \!\! dt  \Bigg[I_e \frac{1}{E_e} \left.
\vphantom{\int_{\int_f}}
\frac{d\sigma}{dx_e} \right|_{\!\!_{x_e=\frac{E_{\gamma'}}{E_e}}} \!\!\!\!\!\!\!\!\!
 e^{-\frac{L_\mathrm{sh}}{l_{\gamma'}}} \left( 1 - e^{-\frac{L_\mathrm{dec}}{l_{\gamma'}}}
\right) \Bigg] \mathrm{BR}_{l\bar{l}}  ,\! \label{eq-Nevents}
\end{equation}
where $N_e$ is the number of the incident electrons, $N_0$ is Avogadro's number, $X_0$ is the unit radiation length of the target material, $L_\mathrm{sh}$ and $L_\mathrm{dec}$ are the lengths of the target plus shield and of the decay region, respectively, and $\mathrm{BR}_{l\bar{l}}$ is the branching ratio of those decay products that the detector is sensitive to, i.e., $e^+e^-$, $\mu^+\mu^-$ or both. The energy distribution $I_e (E_0,E_e,t)$ describes that the initial energy $E_0$ of the electrons in the beam is degraded as they pass through the target. Detailed calculations and expressions are given in~\cite{Bjorken:2009mm,Andreas:2012mt}.

With the partial decay width into leptons given by~\cite{Pospelov:2008zw}
\begin{equation}
\Gamma_{{\gamma'} \rightarrow l^+ l^-} = \frac{\alpha \chi^2}{3} m_{\gamma'}
\left( 1 + 2 \frac{m_l^2}{m_{\gamma'}^2} \right) \sqrt{1 - 4 \frac{m_l^2}{m_{\gamma'}^2}} \ ,
\end{equation} 
the decay length $l_{\gamma'} = \gamma \beta c \tau_{\gamma'}$ can be estimated as
\begin{equation}
l_{\gamma'} ~ \simeq ~ \frac{3 E_{\gamma'}}{\alpha \chi^2 m_{\gamma'}^2} ~  \simeq ~ 8 \ \mathrm{cm} \ 
\frac{E_{\gamma'}}{1 \ \mathrm{GeV}} \ \left(\frac{10^{-4}}{\chi}\right)^2 \ \left(\frac{10 \ \mathrm{MeV}}{m_{\gamma'}}\right)^2. 
\end{equation}
For large values of $\chi$ and/or $m_{\gamma'}$ this is much shorter than the minimum length of the dump required to suppress the SM background, so that electron beam dump experiments can not access the corresponding region in the parameter space.

In an earlier analysis~\cite{Bjorken:2009mm}, limits from the E141 and E137 experiments at SLAC and the Fermilab E774 experiment have been determined. We extended their analysis by two more experiments at KEK in Japan~\cite{Konaka:1986cb} and at the Orsay Linac in France~\cite{Davier:1989wz}. Additionally, in order to derive constraints from Eq.~(\ref{eq-Nevents}), we included the acceptances for the different experiments, taking into account the geometry of the detector and possible energy cuts. For this purpose, we compared the experimental specifications with the events obtained from Monte Carlo simulations with \textsc{MadGraph} for the hidden photon production in Bremsstrahlung followed by the decay, see~\cite{Andreas:2012mt} for details. The limits we obtain for all five electron beam dump experiments are presented in Fig.~\ref{fig-HPbeamdump} (right). A comparison with others constraints is given in~\cite{Andreas:2012mt,Andreas:2011xf:Andreas:2012xh}. An overview of future searches that can further probe the parameter space is presented in~\cite{Hewett:2012ns}.

\section{Hidden sectors with dark matter interacting via hidden photons}\label{sec-DM}

In this section we consider the possibility that the hidden sector also contains DM in addition to the hidden photon. We first discuss the resulting DM relic density and direct detection cross sections in a toy-model and then turn to a more complete supersymmetric realization. The results of this section have been presented in detail in~\cite{Andreas:2011in}.

\subsection{Toy-model: Dirac fermion as dark matter candidate}%

\begin{figure}[b!]
\begin{center}
\includegraphics[height=6cm]{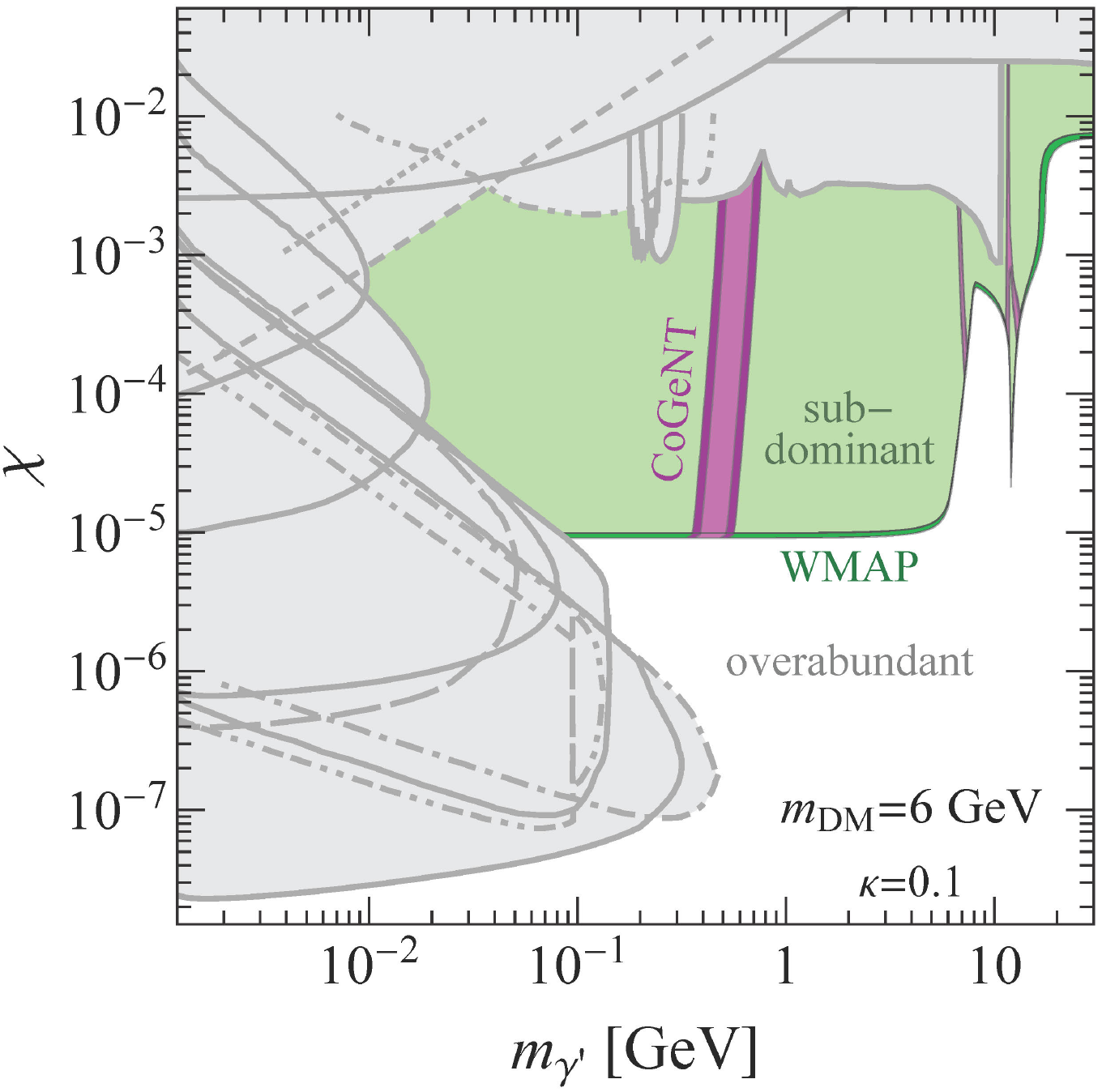}\hspace{0.5cm}
\begin{minipage}{5cm}
\vspace{-5.8cm}\includegraphics[height=3.3cm]{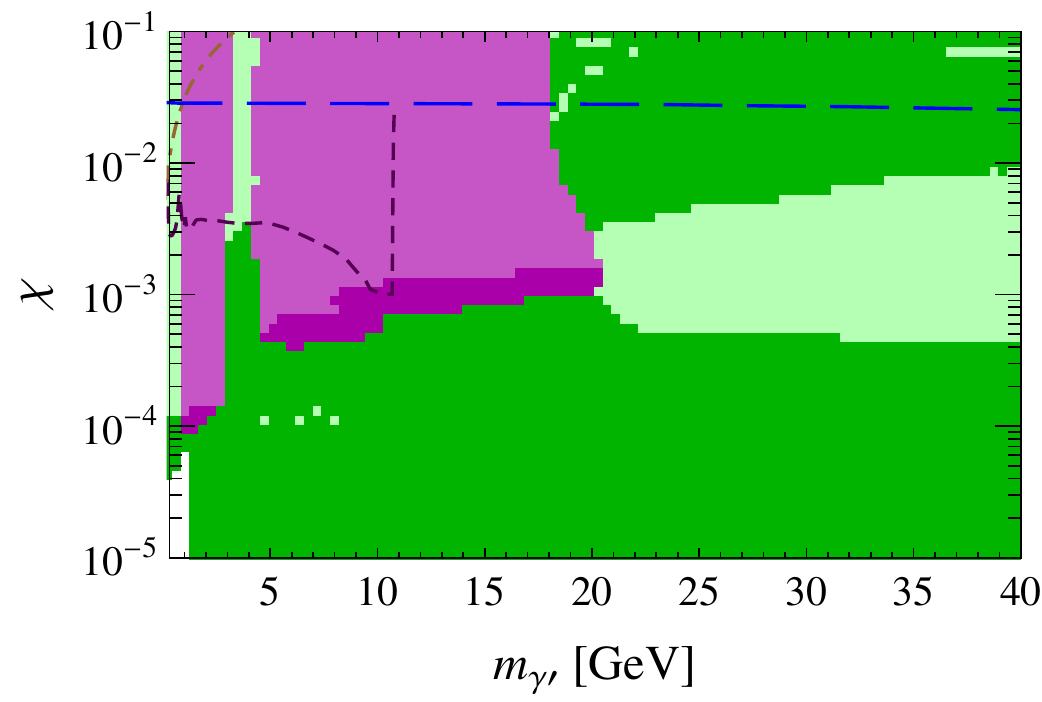}

\vspace{-0.4cm}\includegraphics[height=3.3cm]{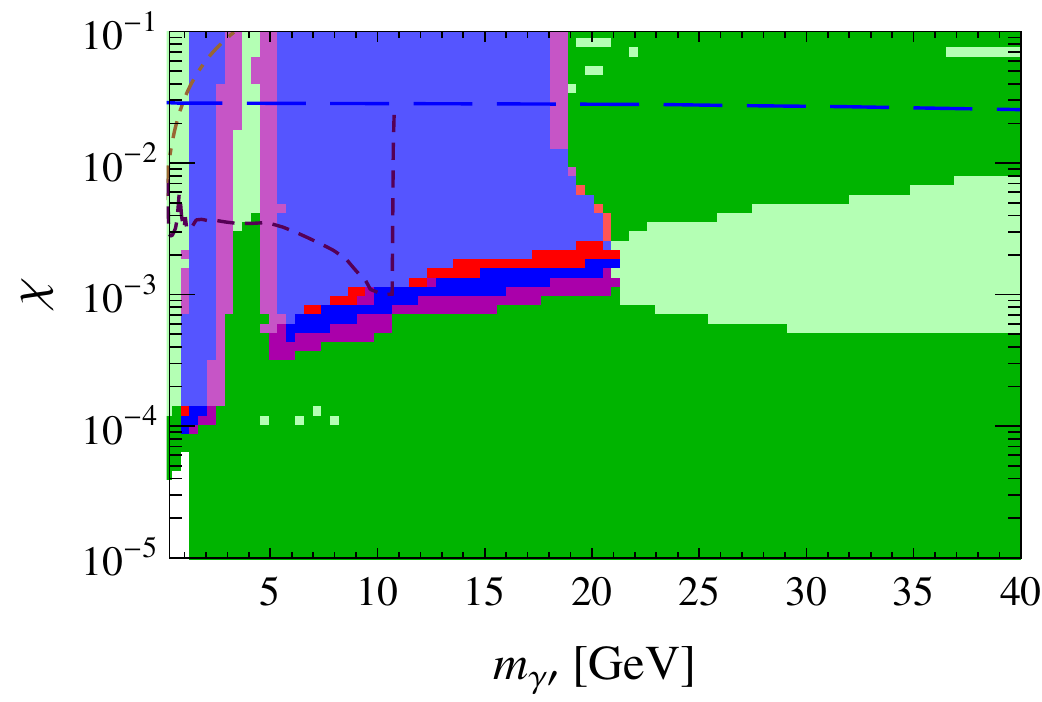}
\end{minipage}\hspace{0.4cm}
\caption{Hidden sector toy-model with Dirac fermion as DM candidate~\cite{Andreas:2011in}. \newline \textit{left:} Viable DM relic abundance (dark green within the WMAP range, light green for subdominant DM) and region where the direct detection rate can explain CoGeNT (purple) for $m_\mathrm{DM} = 6$ GeV, $\kappa=0.1$. Grey areas are excluded by the beam dump experiments of Fig.~\ref{fig-HPbeamdump} (right) and other limits cf.~\cite{Andreas:2012mt,Andreas:2011xf:Andreas:2012xh}.  \newline \textit{right:} Scatter plot scanning over the DM mass for the Standard Halo Model \textit{(top)} and an Einasto profile \textit{(bottom)} for $\kappa=1$. Purple regions are compatible with CoGeNT, red with DAMA, blue with both.} \label{fig-HPDM}
\end{center}
\end{figure}

The simplest possible hidden sector assumed in the following contains despite the hidden photon just one Dirac fermion as DM candidate, cf.~\cite{Pospelov:2007mp:Chun:2010ve:Mambrini:2011dw:Hooper:2012cw} for similar models. Applying the relation given in Eq.~(\ref{eq-kappa}) we fix the hidden sector gauge coupling as a function of the kinetic mixing $\chi$ and determine the DM relic abundance and direct detection rate for fixed $\kappa$. Depending on the masses of both particles the DM annihilation can proceed either in a $s$-channel diagram through the hidden photon into SM particles or in a $t$-channel diagram into two hidden photons. While the former is present for the entire mass range and resonant at $m_{\gamma'} = 2 m_\mathrm{DM}$, the latter is accessible (and dominant) only for $m_\mathrm{DM}>m_{\gamma'}$. For a DM mass of 6 GeV and $\kappa=0.1$, we find that in the dark green band in Fig.~\ref{fig-HPDM} (left) the correct relic abundance can be obtained while in the light green area the contribution to the total DM density is only subdominant. Increasing $\kappa$ pushes the dark green horizontal band upwards to higher values of $\chi$, while it moves down to smaller $\chi$ when $\kappa$ is decreased. The appearance of the resonance at 12 GeV results from the $s$-channel annihilation.  The spin-independent scattering on nuclei is also mediated by the hidden photon and turns out to give cross sections compatible with the ones required to explain the CoGeNT signal for a Standard Halo Model. This is shown in Fig.~\ref{fig-HPDM} (left) as a purple band (90\% CL lighter, 99\% CL darker purple) in which the cross sections for subdominant DM have been rescaled by the actual relic abundance. For a DM mass as light as 6 GeV, there are no constraints from CDMS or XENON. The areas in grey are excluded by the electron beam dump experiments discussed in Sec.~\ref{sec-HP} and various other constraints summarized in~\cite{Andreas:2012mt,Andreas:2011xf:Andreas:2012xh}.

Scanning over the DM mass as a free parameter and keeping $\kappa$ fixed to $\kappa=1$ leads to the scatter plots in Fig.~\ref{fig-HPDM} (right) where the upper one is for a Standard Halo Model and the lower one for an Einasto profile. In both plots, DM in the dark green regions gives the correct relic abundance and in the light green ones a subdominant contribution. The resulting spin-independent scattering cross sections are in agreement with CoGeNT in the purple areas, with DAMA in the red ones and with both experiments in the blue ones (all points shown are in agreement with all other DD limits). For more details and other results please refer to~\cite{Andreas:2011in}. Thus, for a wide range of parameters, the toy-model provides a Dirac fermion as valid DM candidate with the possibility of explaining certain direct detection claims.

\subsection{Supersymmetric model: Majorana and Dirac fermion dark matter}

Embedding the idea of a hidden sector with DM into a more sophisticated and better motivated framework, we construct the simplest anomaly-free supersymmetric model which is possible without adding dimensionful supersymmetric quantities. The corresponding superpotential $W \supset \lambda_S S H_+ H_-$ contains one dimensionless coupling $\lambda_S$ and three chiral superfields $S, H_+, H_-$ where $H_+$ and $H_-$ are charged under the hidden U(1). In the entire analysis, we assume the MSSM in the visible sector, but our results are largely independent of this choice. We consider two mechanisms by which the hidden gauge symmetry is broken and show their different implications on the DM phenomenology. 

\begin{figure}[htb!]
\begin{center}
\includegraphics[width=0.45\textwidth]{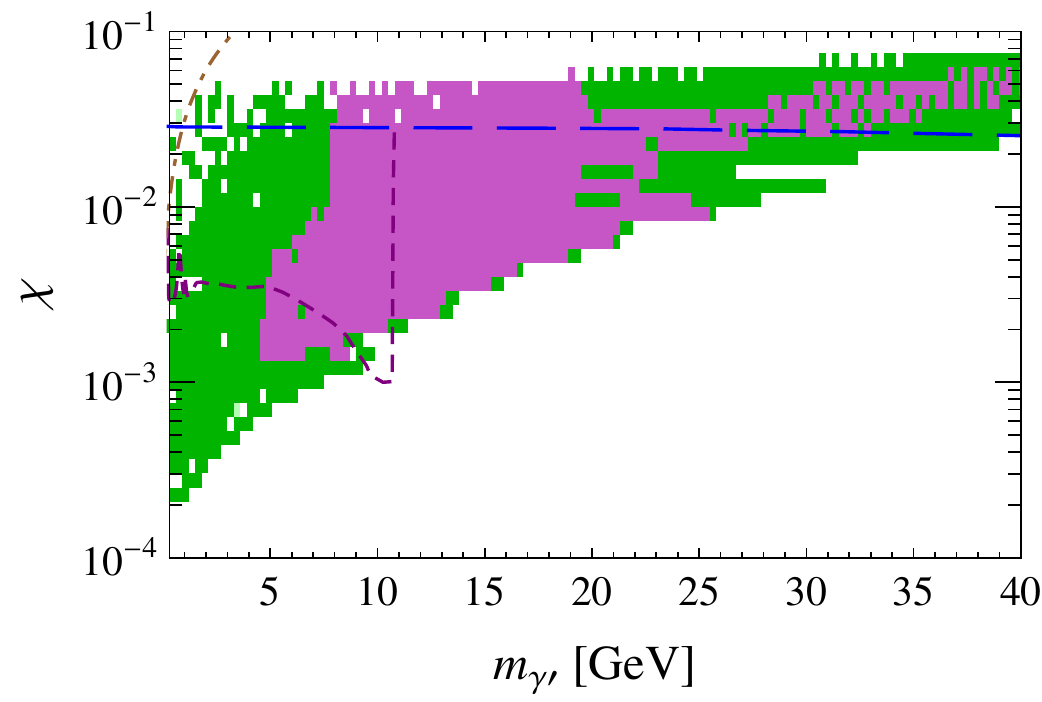}
\includegraphics[width=0.45\textwidth]{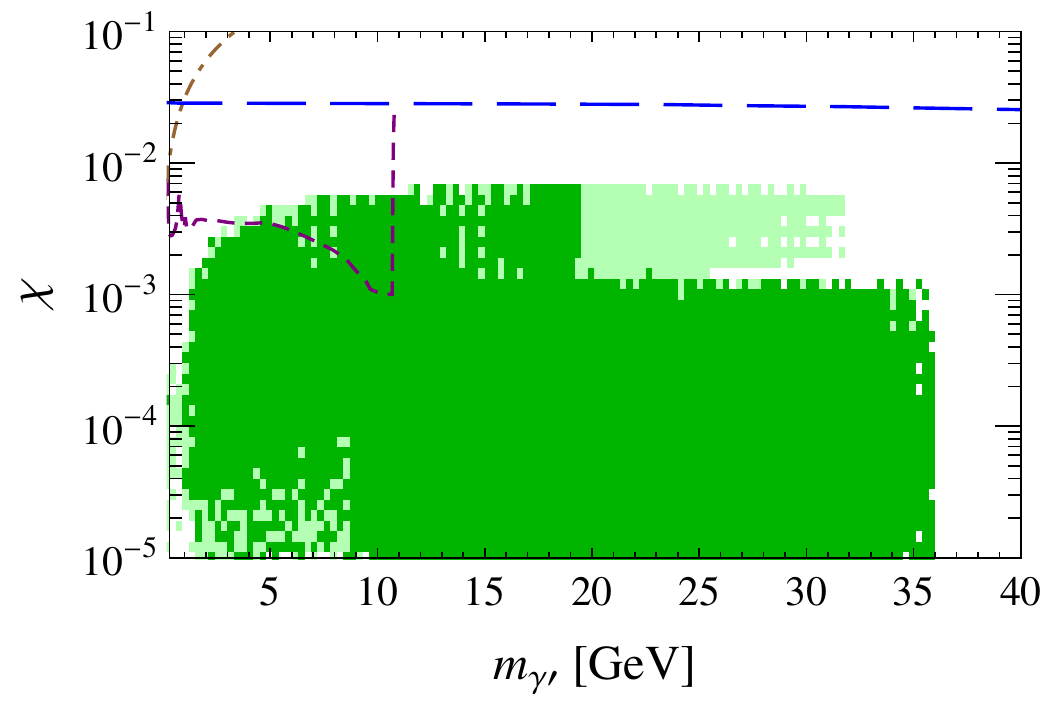}
\caption{Scatter plots for the supersymmetric hidden sector with hidden gauge symmetry breaking induced by the visible sector \textit{(left)} or radiatively \textit{(right)} for $0.1\leq\kappa\leq10$~\cite{Andreas:2011in}. The scattering of the Dirac fermion DM candidate in the left plot is spin-independent while the one of the Majorana fermion in the right plot is mostly spin-dependent. Dark green points give the correct DM relic abundance, light green ones a subdominant contribution and purple ones have spin-independent scattering cross sections in agreement with CoGeNT.} \label{fig-SUSYScattPlot}
\vspace{-0.35cm}
\end{center}
\end{figure}

In the first case, the effective Fayet-Iliopoulos term, which is induced in the hidden sector through kinetic mixing with the visible Higgs D-term, breaks the hidden gauge symmetry. We find that then the DM candidate can be either a Dirac or a Majorana fermion. As in the previous subsection, the Dirac fermion possesses spin-independent scattering on nuclei and thus exhibits a similar phenomenology as the toy-model. This is shown in the scatter plot of Fig.~\ref{fig-SUSYScattPlot} (left) where we scanned both over the DM mass and the parameter $\kappa$ in the range $0.1\leq\kappa\leq10$. Again, the dark green points give the correct relic abundance and for those in purple the direct detection rate is consistent with CoGeNT when a Standard Halo Model is assumed. The plot only contains points for which the scattering cross section is in agreement with constraints from direct detection experiments. In difference to the toy-model, the lower part of the scatter plot can not be filled since the DM particle can not be heavier than the hidden photon and therefore the $t$-channel annihilation is not possible. The Majorana fermion, on the other hand, due to its axial coupling, possesses mostly spin-dependent scattering which is less constrained by direct detection experiments. Spin-independent scattering is possible also for the Majorana fermion thanks to the Higgs-portal term, but the cross sections are tiny, several orders of magnitude below current limits and without any hope of explaining direct detection claims.

In the second case, the running of the Yukawa coupling $\lambda_S$ induces the breaking of the hidden gauge symmetry. The DM candidate turns out to be a Majorana fermion which again, because of its mostly spin-dependent scattering, can not account for the claims in spin-independent direct detection experiments. Scanning over the parameter space and $\kappa$ in the range $0.1\leq\kappa\leq10$ we find points that can give the correct relic abundance or a subdominant contribution as shown in the scatter plot in Fig.~\ref{fig-SUSYScattPlot} (right) in dark and light green, respectively. All points shown are in agreement with the limits arising from spin-dependent direct detection experiments. 

Thus, also supersymmetric hidden sector models can give valid dark matter candidates which, in certain cases, have some similarities to the phenomenology that was obtained in the toy-model.  Results for other parameter settings and plots of the scattering cross sections in the different scenarios compared to experimental limits are given in~\cite{Andreas:2011in}.

\section{Conclusions}
The existence of a hidden sector with a dark force is well motivated from a top-down (string theory, SUSY) and a bottom-up ($g-2$, DM) point of view. Because of the weak interactions with the SM, such scenarios are not much constrained, and we presented here new limits from past electron beam dump experiments on the hidden photon mass and kinetic mixing. If the hidden sector also contains DM, we showed that a toy-model with Dirac fermion DM gives the right relic abundance and spin-independent scattering cross sections able to explain claims in direct detection experiments. For a more sophisticated supersymmetric hidden sector with hidden gauge symmetry breaking induced by the visible sector, we find a Dirac fermion DM candidate with similar phenomenology. A Majorana fermion with mostly spin-dependent scattering can also be the DM candidate in this scenario or when the hidden gauge symmetry is broken radiatively. Our supersymmetric models with gravity mediation have therefore proven to provide viable DM candidates with interesting potential for direct detection experiments.

\section{Acknowledgments}

This work has been done in collaboration with Mark Goodsell, Carsten Niebuhr, and Andreas Ringwald.

\end{document}